\documentclass[apjl]{emulateapj}
\shorttitle{Information of the SDSS Power Spectrum}
\shortauthors{Lee \& Pen}
\begin{document}
\title{Information Content in the Galaxy Angular Power Spectrum from the\\ 
Sloan Digital Sky Survey and Its Implication on Weak Lensing Analysis}
\author{Jounghun Lee}
\affil{Department of Physics and Astronomy, FPRD, Seoul National University, 
Seoul 151-747, Korea : jounghun@astro.snu.ac.kr} 
\author{Ue-Li Pen}
\affil{Canadian Institute for Theoretical Astrophysics, Toronto, ON M5S 3H8, 
Canada: pen@cita.utoronto.ca} 

\begin{abstract}
We analyze the photometric redshift catalog of the Sloan Digital Sky 
Survey Data Release 5 (SDSS DR5) to estimate the Fisher information in the  
galaxy angular power spectrum  with the help of the Rimes-Hamilton technique. 
It is found that the amount of Fisher information contained in the galaxy 
angular power spectrum is saturated at lensing multipole scale  
$300\le l\le 2000$ in the redshift range $0.1\le$photo-z$<0.5$. 
At $l=2000$, the observed information is two orders of magnitude lower than 
the case of Gaussian fluctuations. This supports observationally that the 
translinear regime of the density power spectrum contains little independent 
information about the initial cosmological conditions, which is consistent 
with the numerical trend shown by Rimes-Hamilton. Our results also suggest 
that the Gaussian-noise description may not be valid in weak lensing 
measurements.
\end{abstract}
\keywords{cosmology:theory --- methods:statistical --- 
large-scale structure of universe}

\section{INTRODUCTION}

Cosmic shear, which refers to the weak gravitational lensing by large-scale 
structure in the universe, is an invaluable tool in cosmology since it 
directly probes the gravitational clustering of dark matter in the universe 
without concerns about light-to-matter bias.  Very recently, the three 
dimensional map of dark matter distribution has been indeed reconstructed 
on cosmological scales from the measurements of cosmic shear \citep{Massey07}, 
which marks an advent of {\it lensing cosmology}.  Among many lensing 
observables, the angular power spectrum is regarded particularly important, 
since it is in principle capable of constraining the key cosmological 
parameters with high precision, including the enigmatic nature of 
dark energy \citep[see][for a recent review]{HJ08}. 

The success of the angular power spectrum as a cosmological probe, however, is 
subject to one critical issue: how much information does it preserve about 
the initial conditions of the universe?  A natural expectation is that some 
but not all information might have been destroyed in the subsequent nonlinear 
evolution. The amount of information contained in the power spectrum about 
the initial cosmological conditions is directly related to the precision of 
the cosmological parameters constrained by using the matter power spectrum. 
In other words, the non-Gaussian errors caused by the loss of information 
in the power spectrum would propagate into large uncertainties in the 
determination of the cosmological parameters. The error propagation and 
precision on cosmological parameters can be quantified in terms of the 
Fisher information contained in the matter power spectrum \citep{TTH97}. 

\citet[][hereafter, RH05]{RH05} have for the first time 
estimated the Fisher information in the matter power spectrum.
Basically, RH05 have calculated the information content about the amplitude 
of the linear power spectrum averaged over an ensemble of many N-body 
realizations and found very little independent information at the translinear 
scale. Somewhat surprisingly, however, RH05 also found that there is a sharp 
rise in the amount of information in the nonlinear scale 
\citep[see also][]{RH06,RHS06}. This phenomenon has been also noted 
by the analytic work of \citet{NSR06}.

Yet, according to the recent results derived by \citet{NS07} based on the 
halo model, the Fisher information in the density power spectrum about 
all key cosmological parameters including the initial amplitude is highly 
degenerate both in the translinear and the nonlinear regime. Their work 
has indicated that it might not be possible to extract the initial 
cosmological conditions from the nonlinear dark matter power spectrum 
to a high statistical accuracy.

Thus, the previous numerical and analytic results forecast that the 
accuracy in the determination of the cosmological parameters in lensing 
cosmology may be lower due to the non-Gaussian errors.  
As the next generation of large surveys optimized for weak lensing will soon 
be on the pipeline, it is imperative to test observationally the information 
content in the lensing power spectrum. In this Letter we attempt to do this 
test by applying the RH05 technique to the photometric galaxy catalogs from 
the Sloan Digital Sky Survey Data Release 5 \citep[SDSS DR5,][]{AM07} 
at typical lens redshifts $z\sim 0.3$.

\section{DATA}

We use the {\tt Photoz2} and the {\tt PhotoPrimary} catalog which are both 
publicly available at the web site of the SDSS DR5 
({\tt http://www.sdss.org/dr5}).  The {\tt Photoz2}-catalog contains 
information on galaxy photometric redshift (photo-z, $z_{p}$), while the 
{\tt PhotoPrimary} catalog has imaging data of galaxy right ascension 
($\alpha$), declination ($\delta_{I}$), and dereddened model magnitude 
($M_{r}$). For our analysis, we select those galaxies in the catalog whose 
dereddened model magnitude, photo-z and angular positions satisfy the 
constraints of $M_{r}<22.0$; $0.0\le z_{p}<1$; 
$0^{\circ}\le \delta_{I} < 60^{\circ}$; 
$120^{\circ}\le \alpha < 240^{\circ}$. 

Selected are a total of $42587179$ galaxies which we divide into smaller 
samples according to their angular positions and photo-z's. We first decompose 
the selected portion of the sky into 55 cells, each having approximately same 
linear size of $10^{\circ}$. To make equal-size cells on {\it non-flat} sky, 
the bin size of $\delta_{I}$ is set at the constant value of 
$\delta_{I}=10^{\circ}$, while the bin size of $\alpha$ is adjusted by a 
factor of $1/\cos\delta_{Im}$ 
to compensate for the decrease in the width of $\Delta\alpha$ with 
$\delta_{I}$. Here, we choose $\delta_{Im}$ as the median value of 
$\delta_{I}$ at each cell. We end up with having a total of $55$ 
cells on the plane of the sky in the selected angular range. 
Table \ref{tab:data} lists the ranges of $\alpha$ and $\delta_{I}$, a total 
number of galaxies ($N_{g}$) and cells ($N_{C}$) in four photo-z slices.

We treat approximately each cell as a square region and decompose it further 
into $128^{2}$ pixels of equal size. Counting the number of galaxies 
at each pixel, we find those pixels where no galaxy is located and mask them 
and their nearest neighbor pixels  as gap regions, assigning zero density 
contrast. After masking the gap regions, we bin the selected photo-z range 
with bin size of $\Delta z_{p}=0.1$. A set of $55$ cells in each redshift 
slice represents an ensemble of realizations for the calculation of the 
galaxy angular power spectrum. The size of the field is well in linear 
regime, so the small-size cells should be reasonably independent. Note that 
we don't include the effect of beat-coupling discussed in \citet{RH06} and 
in \citet{RHS06}.

For each cell in each photo-z slice, we construct the two dimensional density 
field on $128^{2}$ pixels by determining the dimensionless density contrast 
$\delta$ as residual number density of the galaxies, 
$\delta\equiv (n-\bar{n})/\bar{n}$,  where $n$ is the number density of the 
galaxies belonging to a given pixel, and $\bar{n}$ is the mean number density 
of galaxies averaged over a given cell in a given photo-z slice. 
For the calculation of the mean number density $\bar{n}$, we exclude the 
gap regions, counting only unmasked pixels.

\begin{deluxetable}{ccccc}
\tablewidth{0pt}
\setlength{\tabcolsep}{5mm}
\tablehead{photo-z &  $\delta_{I}$ & $\alpha$ & \# of galaxies & \# of cells} 
\startdata
$[0.1,0.2)$ & $[0^{\circ},60^{\circ})$ & 
$[120^{\circ}, 240^{\circ})$ & $2794590$ & $55$ \\ 
$[0.2,0.3)$ & $[0^{\circ},60^{\circ})$ 
& $[120^{\circ}, 240^{\circ})$ & $4865953$ & $55$ \\ 
$[0.3,0.4)$ & $[0^{\circ},60^{\circ})$ 
& $[120^{\circ}, 240^{\circ})$ & $9768987$ & $55$ \\
$[0.4,0.5)$ & $[0^{\circ},60^{\circ})$ 
& $[120^{\circ}, 240^{\circ})$ & $9500664$ & $55$ \\
\enddata
\label{tab:data}
\end{deluxetable}

Then, we perform the Fourier-transformation of the two dimensional density 
field with the help of the FFT routine provided in the numerical recipes 
\citep{Press92} and calculate the galaxy angular power spectrum, $C_{G}(l)$, 
as a function of a multipole $l$ in the Fourier space:
\begin{equation}
\label{eqn:wg}
C_{G}(l) = \frac{N_{\rm total}}{N_{\rm unmask}}
\langle\vert\tilde{\delta}({\bf l})\vert^{2}\rangle,
\end{equation}
where ${\bf l}$ is a two dimensional multipole vector representing the Fourier 
transform of the two dimensional angular position vector, $N_{\rm total}$ is 
the total number of pixels in a given cell (i.e., $N_{\rm total}=128^{2}$), 
and $N_{\rm unmask}$ represents the number of pixels which do not belong to 
the masked gap regions in a given cell. We weight the angular power spectrum 
by a factor of $(N_{\rm unmask}/N_{\rm total})^{-1}$ to compensate for the 
zero powers assigned to the gap regions \citep{Hamilton05}. The average 
fraction of $N_{\rm unmask}/N_{\rm total}$ is found to be $0.92\pm 0.02$.

The galaxy angular power spectrum, $C_{G}(l)$, obtained by equation 
(\ref{eqn:wg}) includes the shot-noise. To eliminate the shot-noise power 
from a given cell to which a total of $N_{g}$ galaxies belong, we construct 
a random sample of equal number of galaxies by generating the values of 
($\delta_{I}$, $\alpha$) randomly. If a randomly generated galaxy is found 
to fall in gap regions, we regenerate ($\delta_{I}$,$\alpha$) till it 
belongs to a unmasked pixel. Using this random sample of $N_{g}$ galaxies, 
we repeat the whole process: constructing the density field on 
$128^{2}$ pixels, Fourier-transforming the density field, calculating 
the shot-noise power spectrum, $C_{\rm SN}(l)$,  which is also weighted 
by the same factor of $N_{\rm total}/N_{\rm unmask}$. 
Finally, we subtract the shot-noise power spectrum, $C_{SN}$, from 
$C_{G}(l)$, to obtain the shot-noise-free angular power spectrum, 
$C(l)$ for all $55$ cells in each photo-z slice.
\begin{figure}
\begin{center}
\includegraphics[width=3.2in]{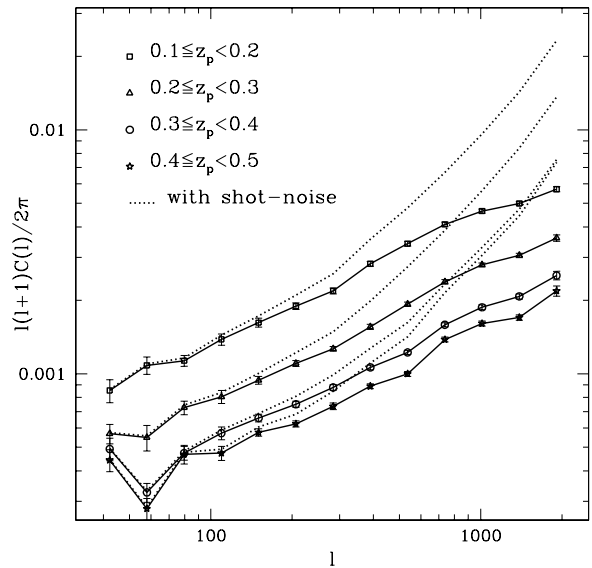}
\caption{Mean angular power spectrum in a dimensionless form, 
$C(l)l(l+1)/2\pi$, averaged over $55$ cells in four photo-z slices: 
$0.1\le z<0.2$ (open squares); $0.2\le z<0.3$ (open triangles); 
$0.3\le z<0.4$ (open circles); $0.4\le z<0.5$ (open stars). 
The errors are calculated as one standard deviation in the measurement 
of the mean angular power spectrum. The dotted lines represent the cases 
before the shot-noise subtraction.}
\label{fig:pow}
 \end{center}
\end{figure}

Figure \ref{fig:pow} plots the mean angular power spectra, 
$l(l+1)C(l)/2\pi$, averaged over $55$ cells after the shot-noise 
subtraction as solid lines with errors for four different photo-z slices. 
The errors, $\sigma_{C}$, are calculated as one standard deviation in the 
measurement of the mean angular power spectrum averaged over $55$ cells 
belonging to each photo-z slice: 
$\sigma_{C}\equiv \sqrt{\langle\Delta C^{2}(l)\rangle/(N_{C}-1)}$. 
For the comparison, the mean angular power spectrum, $\bar{C}_{G}(l)$, before 
the shot-noise subtraction is also plotted as dotted lines.
As can be seen, before the shot-noise subtraction, the slope of the mean 
angular power spectrum changes from $\sim 1$ to $\sim 2$ as the multipole 
$l$ increases. After the shot-noise subtraction, its slope stays constant 
at $\sim 1$. 

The fluctuations of the angular power spectrum at the lowest multipole 
($l\le 100$) represent the presence of the sample variance caused by the 
finite size of the sky cells. Note that the amplitude of the angular power 
spectrum decreases with photo-z, indicating the growth of powers.  
Here, we focus mainly on the photo-z ranges of $[0.1,0.5)$: 
At higher photo-z ($z_{p}\ge 0.5$), the shot-noise will dominate; 
At lower photo-z ($z_{p}<0.1$), due to the increase of the fractional 
distance errors, the three dimensional analysis will be necessary 
to measure the power spectrum \citep{Dodelson01,Tegmark02}.

\section{ANALYSIS}

The angular power spectrum estimated in \S 2 is supposed to contain  
information about the amplitude of the linear power spectrum:
$C(l)=C(l;\ln A)$ where $\ln A$ is the log of the amplitude of the linear 
density power spectrum. According to RH05, the Fisher information 
contained in the angular power spectrum about the amplitude of the linear 
power spectrum can be written as \citep{TTH97}
\begin{equation}
\label{eqn:I}
I \equiv -\bigg{\langle}\sum_{i,j}
\frac{\partial\ln C(l_{i})}{\partial\ln A}{\rm Cov}^{-1}_{ij}
\frac{\partial\ln C(l_{j})}{\partial\ln A}\bigg{\rangle}.
\end{equation}
Here $({\rm Cov}^{-1}_{ij})$ represents the inverse of the covariance matrix 
whose components are given as 
\begin{equation}
\label{eqn:cov}
{\rm Cov}_{ij} = \frac{\langle\Delta C(l_{i})\Delta C(l_{j})\rangle}
{\bar{C}(l_{i})\bar{C}(l_{j})},
\end{equation}
where $\Delta C(l)\equiv C(l)-\bar{C}(l)$ is the scatter of the angular power 
spectrum from the mean value at a given multipole $l$. As explained in detail 
by \citet{RH06}, for the case of linear power spectrum, $C^{0}(l)$, 
from the Gaussian density fluctuations, the inverse of the covariance matrix 
is diagonal and equation (\ref{eqn:cov}) will be simplified into
 $I(l_{i}) = n(l_{i})/2$ since 
$\partial\ln C(l)/\partial\ln A = 1$ and ${\rm Cov}^{-1}_{ii}=n(l_{i})/2$ 
where $n(l_{i})$ is the number of Gaussian modes around the 
multipole $l_{i}$. 

RH05 asserted that the amount of information preserved in the nonlinear power 
spectrum depends on the existence of an invertible one-to-one mapping between 
the linear and nonlinear regime. If such a mapping exists, then the same 
amount of information that the linear power spectrum contains about the 
initial amplitude can be extracted also from the nonlinear power spectrum. 
If not, then it would be difficult to determine the initial amplitudes with 
high precision from the nonlinear power spectrum.  However, note that what 
we have measured in \S 2 is the {\it nonlinear galaxy power spectrum}. 
Unlike the nonlinear matter power spectrum that is reasonably well understood 
\citep{HA91,PD96}, the non-linear clustering of galaxies is rather 
heuristic and usually constructed to fit the data. Although the non-linear 
galaxy power spectrum $C(l)$ is known to be neither the nonlinear matter 
power spectrum nor the linear matter power spectrum, we set $C(l)$ to the 
linear matter power spectrum $AC^{0}(l)$ for simplicity.
\begin{figure*}
\begin{center}
\includegraphics[width=0.75\textwidth]{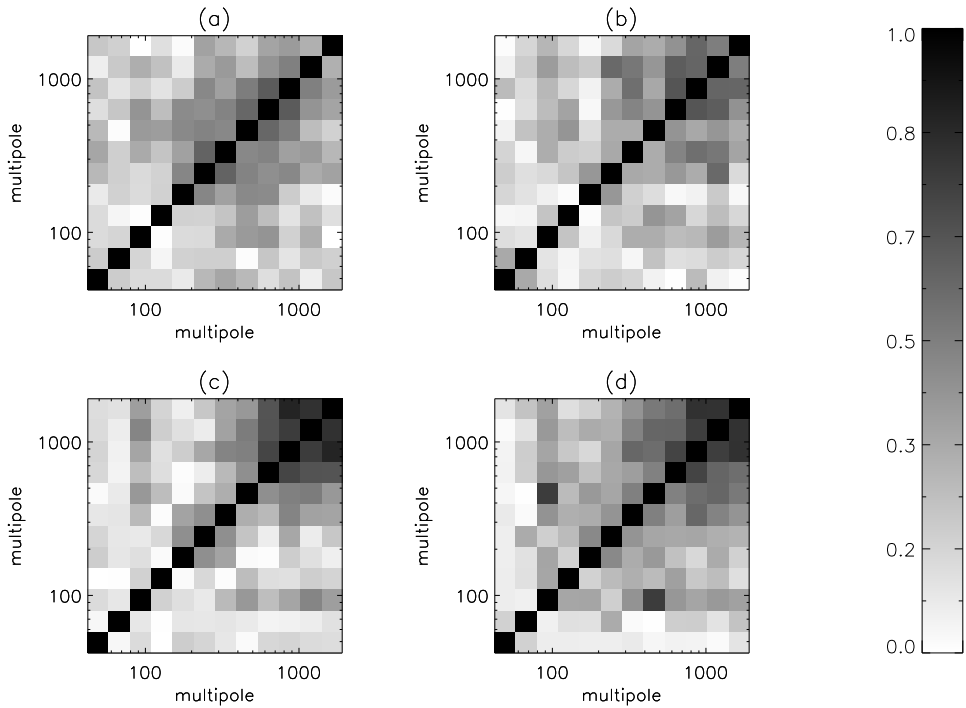}
\caption{Covariance matrices of the galaxy angular power spectrum, $C(l)$, 
at four photo-z slices: (a) $0.1\le z_{p}<0.2$, (b) $0.2\le z<0.3$, 
(c) $0.3\le z_{p}<0.4$ and (d) $0.4\le z_{p}<0.5$. As done in \citet{RH05}, 
the grey scale is used to represent the magnitude of the covariances.}
\label{fig:inf}
\end{center}
\end{figure*}

\begin{figure}
\begin{center}
\includegraphics[width=3.2in]{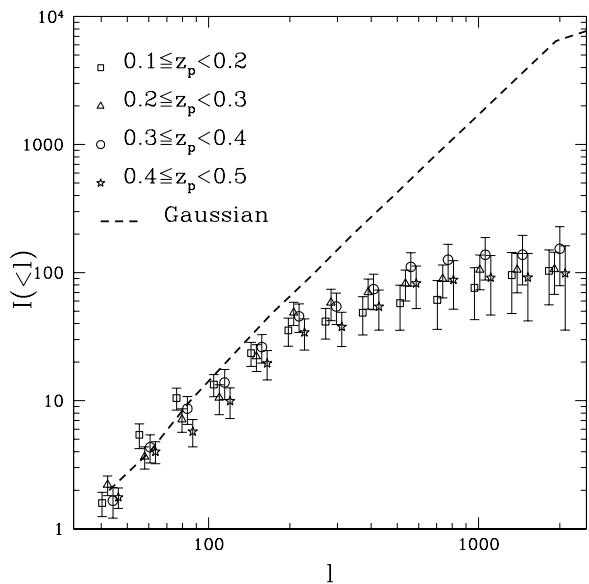}
\caption{Cumulative information in the angular power spectrum for the four 
photo-z ($z_{p}$) slices. The dashed line represents the cumulative 
information in the linear angular power spectrum and the errors are 
calculated from 1000 times bootstrap-resampling.}
\label{fig:cov}
\end{center}
\end{figure}

Using the mean angular power spectra measured at four photo-z slices, 
we first calculate the normalized covariance matrix defined as  
\begin{equation}
\hat{\rm Cov}_{ij} = \frac{\langle\Delta C(l_{i})\Delta C(l_{j})\rangle}
{\sqrt{\langle\Delta C^{2}(l_{i})\rangle\langle\Delta C^{2}(l_{j})\rangle}}.
\end{equation}
Figure \ref{fig:cov} plots the magnitudes of ($\hat{\rm Cov}_{ij}$) as grey 
scales for the cases of four photo-z slices: $[0.1,0.2)$, $[0.2,0.3)$, 
$[0.3,0.4)$ and $[0.4,0.5)$ in the panel (a), (b), (c) and (d), respectively. 
As can be seen, the magnitudes of the off-diagonal elements of 
($\hat{\rm Cov}_{ij}$) deviate from zero, which suggests the reduced 
amount of information than the case of Gaussian fluctuations.

Finding the inverse of the covariance matrix, ${\rm Cov}^{-1}$, we 
calculate the cumulative information $I(<l)$ contained in the SDSS 
angular power spectrum: 
\begin{equation}
I(<l_{k}) = \sum_{j=1}^{k}\sum_{i=1}^{k}y_{i}{\rm Cov}^{-1}_{ij}y_{j}, 
\qquad k=1,\ldots, n_{l}
\end{equation}
where ${\bf y}\equiv (y_{i})$ is a column vector whose component is 
all unity and $n_{l}$ is the total number of multipole bins. 
Figure \ref{fig:inf} plots $I(<l)$ for the cases of four photo-z slices. 
The errors are calculated from the 1000 times bootstrap resampling, and 
the dashed line represents the case of Gaussian-fluctuations. 
We offset slightly the horizontal positions of the four symbols at each 
multipole bin to show the errors clearly. 

If there were a one-to-one map from the nonlinear to the linear regimes, 
then $I(<l)$ obtained from SDSS catalogs would follow the dashed line. 
As can be seen, at low multipoles ($l<300$) it matches the 
Gaussian case pretty well. At higher multipole ($l\ge 300$), however, we 
detect a clear signal of saturation in $I(<l)$. Note that at $l=2000$ the 
observed amount of the cumulative information is two orders of magnitude 
lower than expected for the case of Gaussian-fluctuations in all four 
photo-z slices.  The amplitude of the saturated region of $I(<l)$ at high 
$l$ tends to increase as photo-z increases in the range of 
$0.1\le z_{p}<0.4$, which is consistent with the numerical results of RH05.  
Although the amplitude becomes lower at the highest photo-z slice 
($0.4\le z_{p}<0.5$), the differences are within the errors. 
Unlike RH05, however, we could not find a sharp rise of $I(<l)$ at nonlinear 
scales even when we increase the number of pixels.  It might be because in 
the nonlinear scale the light-to-matter bias becomes important and many 
other complicated baryonic processes might have destroyed the information 
content further. 

\section{Discussion and Conclusion}

It is worth mentioning here that what you have calculated using the SDSS data 
is not rigorously the information about the amplitude of the power spectrum.
But, it only approximates the information in the linear power spectrum to 
the extent that $d\ln P/d\ln A$ is near unity.

Our result provides an observational evidence for the previous theoretical 
clues that at translinear scale there is very little independent information 
in the matter power spectrum, consistent with light tracing matter till this 
scale \citep{RH05,RH06,RHS06, NSR06,NS07}. Since the multipole scales where 
the information saturation is detected correspond to the weak lensing regime, 
this result has a direct impact on the weak lensing analyses. The loss of 
information in the angular power spectrum would lead to increasing 
non-Gaussianity in sampling errors. Unlike the usual assumption adopted in 
most weak lensing analyses that the non-Gaussianity contribution to the 
sampling errors for the angular power spectrum is marginal 
\citep[][and references therein]{WH00,CH01}, our result suggests that it 
should be quite substantial and thus the Gaussian-noise description for the 
lensing power spectrum should not be valid. 

\acknowledgments

We thank our referee, A. Hamilton for useful suggestions. 
We also thank K. Kratter and S. Bonoli for their many helps with the IDL 
routines. J.L. acknowledges the financial support from the Korea Science 
and Engineering Foundation (KOSEF) grant funded by the Korean Government 
(MOST, NO. R01-2007-000-10246-0).

Funding for the SDSS and SDSS-II has been provided by the Alfred
P. Sloan Foundation, the Participating Institutions, the National
Science Foundation, the U.S. Department of Energy, the National
Aeronautics and Space Administration, the Japanese Monbukagakusho, the
Max Planck Society, and the Higher Education Funding Council for
England. The SDSS Web Site is http://www.sdss.org/. 

The SDSS is managed by the Astrophysical Research Consortium for the
Participating Institutions. The Participating Institutions are the
American Museum of Natural History, Astrophysical Institute Potsdam,
University of Basel, University of Cambridge, Case Western Reserve
University, University of Chicago, Drexel University, Fermilab, the
Institute for Advanced Study, the Japan Participation Group, Johns
Hopkins University, the Joint Institute for Nuclear Astrophysics, the
Kavli Institute for Particle Astrophysics and Cosmology, the Korean
Scientist Group, the Chinese Academy of Sciences (LAMOST), Los Alamos
National Laboratory, the Max-Planck-Institute for Astronomy (MPIA),
the Max-Planck-Institute for Astrophysics (MPA), New Mexico State
University, Ohio State University, University of Pittsburgh,
University of Portsmouth, Princeton University, the United States
Naval Observatory, and the University of Washington.


\clearpage
\begin{thebibliography}{}

\bibitem[Adelman-McCarthy et al.(2007)]{AM07}
{Adelman-McCarthy}, J.~K.,  et al. 2007, \apj, 172, 634

\bibitem[Cooray \& Hu(2001)]{CH01}
{Cooray}, A. \& Hu, W. 2001, \apj, 554, 56

\bibitem[Dodelson et~al.(2001)]{Dodelson01}
{Dodelson}, S., et al. 2001, \apj, 572, 140

\bibitem[Hamilton et~al.(1991)]{HA91}
{Hamilton}, A.~J.~S., {Kumar}, P., {Lu}, E. \& {Matthews}, A. 1991, 
\apj, 374, L1

\bibitem[Hamilton (2005)]{Hamilton05}
{Hamilton}, A.~J.~S. 2005, Data Analysis in Cosmology proceedings of an 
International Summer School, (Valencia: Springer-Verlag) 

\bibitem[Hamilton et~al.(2006)]{HRS06}
{Hamilton}, A.~J.~S., {Rimes}, C. D., \& {Scoccimarro}, R. 2006, 
\mnras, 371, 1188

\bibitem[Hoekstra \& Jain(2008)]{HJ08}
{Hoekstra}, H. \& {Jain}, B. 2008, Ann. Rev. Nuclear and Particle Science, 
58, 1056

\bibitem[Massey et al.(2007)]{Massey07}
Massey, R., et al. 2007, Nature, 445, 286

\bibitem[Neyrinck et al.(2006)]{NSR06}
Neyrinck, M.~C., Szapudi, I. \& Rimes, C. D. 2006, \mnras, 370, L66

\bibitem[Neyrinck \& Szapudi(2007)]{NS07}
Neyrinck, M.~C. \& Szapudi, I. 2007, \mnras, 375, L51

\bibitem[Rimes \& Hamilton(2005)]{RH05}
{Rimes}, C.~D., \& {Hamilton}, A.~J. S. 2005, \mnras, 360, L82

\bibitem[Rimes \& Hamilton(2006)]{RH06}
{Rimes}, C.~D., \& {Hamilton}, A.~J. S. 2006, \mnras, 371, 1205

\bibitem[Rimes et al.(2006)]{RHS06}
{Rimes}, C.~D., {Hamilton}, A.~J. S., \& {Scoccimarro}, R. 2006, 
\mnras, 371, 1188

\bibitem[Peacock \& Dodds(1996)]{PD96}
Peacock, J.~A. \& Dodds, S.~J. 1996, \mnras, 280, L19

\bibitem[Press et al.(1992)]{Press92}
{Press},  W.~H.,  et al. 1992, Numerical Recipes in Fortran 
(Cambridge: Cambridge Univ. Press)

\bibitem[Tegmark et al.(1997)]{TTH97}
{Tegmark}, M., {Taylor}, A.~N., \& {Heavens}, A. F. 1997, 
\apj, 480, 22

\bibitem[Tegmark et al.(2002)]{Tegmark02}
{Tegmark}, M., et al. 2002, \apj, 571, 191

\bibitem[White \& Hu(2000)]{WH00}
White, M. \& Hu W. 2000, \apj, 537, 1

\end{thebibliography}
\end{document}